\documentclass[11pt,a4paper]{article}

\usepackage{amssymb}
\usepackage{amsmath}
\usepackage{autobreak}
\usepackage{graphicx}
\usepackage{bm}
\usepackage{mathrsfs}
\usepackage{cite}
\usepackage{color}

\begin{document}

\title{The four-loop $\beta$-function from vacuum supergraphs and the NSVZ relation for ${\cal N}=1$ SQED regularized by higher derivatives}

\author{I.E.Shirokov, V.Yu.Shirokova\, $\vphantom{\Big(}$
\\
{\small{\em Moscow State University, }}\\
{\small{\em Faculty of Physics, Department of Theoretical Physics,}}\\
\medskip
{\small{\em 119991, Moscow, Russia.}}
}

\maketitle

\begin{abstract}
In ${\cal N}=1$ SQED with $N_f$ flavors regularized by higher derivatives we obtain the four-loop beta function using a method based on calculating vacuum supergraphs. For this purpose we use a special C++ program which obtain contributions to the $\beta$-function from supergraphs without external legs in the form of integrals of double total derivatives. After that the result was compared with the three-loop anomalous dimension calculated earlier. We explicitly check the NSVZ relation in this order.
\end{abstract}

\allowdisplaybreaks

\section{Introduction}
\hspace*{\parindent}
It is known that supersymmetric theories have a better ultraviolet behaviour in comparison with the non-supersymmetric ones \cite{West:1990tg,Gates:1983nr, Wess:1992cp, Buchbinder:1998qv}. This fact manifests itself in the form of the existence of non-renormalization theorems. In addition to well-known statements such as the non-renormalization of superpotential \cite{Grisaru:1979wc}, there is another interesting one. This is the so-called NSVZ equation \cite{Novikov:1983uc,Jones:1983ip,Novikov:1985rd,Shifman:1986zi}, which relates the $\beta$-function to the anomalous dimension of the matter superfields in previous orders. This equation was found rather long ago, but until recently there was no all-loop perturbative proof of it. Most perturbative calculations in the supersymmetric case were made with the regularization by dimensional reduction \cite{Siegel:1979wq}. However, this regularization is mathematically inconsistent \cite{Siegel:1980qs,Jack:1997sr}. As a consequence, it can break supersymmetry in
higher loops \cite{Avdeev:1981vf,Avdeev:1982xy,Avdeev:1982np,Velizhanin:2008rw}. Several calculations made in $\overline{DR}$-scheme  \cite{Avdeev:1981ew, Jack:1996vg,Jack:1996cn,Harlander:2006xq} (where the regularization by dimensional reduction is supplemented by modified
minimal subtraction \cite{Bardeen:1978yd}) showed that the NSVZ relation is valid only in the lowest loops where the scheme dependence of the renormalization group functions is not significant. In order for it to remain valid in higher loops, one needs to perform a specially constructed finite renormalization in each order of the perturbation theory \cite{Jack:1996vg,Jack:1996cn,Jack:1998uj,Harlander:2006xq,Mihaila:2013wma}.

A much more convenient regularization is the higher covariant derivatives method \cite{Slavnov:1971aw,Slavnov:1972sq}. This regularization can easily be adopted for the supersymmetric case  \cite{Krivoshchekov:1978xg, West:1985jx}. It turned out that for renormalization group functions defined in terms of bare coupling constants, when using regularization by higher covariant derivatives, the NSVZ relation is valid in all loops, both in the Abelian \cite{Stepanyantz:2011jy,Stepanyantz:2014ima} and non-Abelian case \cite{Stepanyantz:2020uke}. In addition, for renormalization group functions defined in terms of a renormalized coupling constant, there is an all-loop renormalization prescription called HD+MSL (higher derivatives regularization plus minimal subtractions of logarithms) prescription for which this relation also holds \cite{Kataev:2013eta, Shakhmanov:2017wji,Stepanyantz:2017sqg }. These results exactly agree with the explicit three-loop calculations made in \cite{Kataev:2013eta,Stepanyantz:2012zz, Stepanyantz:2012us, Kataev:2013csa,Kataev:2014gxa}.

 It should also be mentioned that in the Abelian case the NSVZ relation arises due to the fact that in the massive theory all higher-loop divergences in the gauge part of the effective action appear from the mass renormalization in the one-loop term $\ln \Lambda / m_{0}$, see \cite{Shifman:1986zi, Smilga:2004zr} for details. Consequently, the higher-order contributions to the $\beta$-function are related to the anomalous dimension. In particular, the NSVZ equation is satisfied for renormalization group functions defined in terms of the renormalized coupling constant in the on-shell scheme in all orders \cite{Kataev:2019olb}. 

Note that multiloop calculations are certainly very important to study some specific properties of the quantum corrections and can reveal some scheme depended properties. For example, in  \cite{Shirokov:2022jyd} the three-loop calculation of the anomalous dimension revealed the existence of the so-called minimal scheme where all scheme depended terms can be set to zero. The three-loop results were used to check the general explicit expressions for arbitrary coefficients at powers of  logarithms constructed in \cite{Meshcheriakov:2022tyi} (see also \cite{Meshcheriakov:2023fmk}). However, the multiloop calculations made with the help of the higher covariant derivative regularization (see, e.g., \cite{Pimenov:2009hv, Aleshin:2016rrr, Shakhmanov:2017soc, Kazantsev:2018nbl, Kazantsev:2018kjx, Kuzmichev:2019ywn, Aleshin:2020gec, Kazantsev:2020kfl}) are very complicated even in the two- and three-loop approximation. For making such calculations I.S. wrote a new C++ program \cite{Shirokov:2022qdw}. It can process the algebra of supersymmetric covariate derivatives and generate all the supergraphs in the given order. Until now it could only work with two point Green functions of matter superfields, needed for calculating the anomalous dimension.

Nevertheless, the calculation of the $\beta$-function is also of great importance. Usually, to obtain it, one needs to calculate diagrams with two external gauge field lines. In supersymmetric gauge theories regularized by higher derivatives it was shown that integrals giving the $\beta$-function are integrals of total derivatives  \cite{Soloshenko:2003nc} and double total derivatives \cite{Smilga:2004zr}. Using this fact a new method of calculation was proposed \cite{Stepanyantz:2019ihw}. It is based on calculating specially modified vacuum supergraphs. The result obtained using this method is expressed in the form of integrals of double total derivatives. This method has already been used in several calculations  \cite{Stepanyantz:2019ihw,Kuzmichev:2019ywn,Stepanyantz:2019lyo, Aleshin:2020gec, Aleshin:2022zln}. However, even despite the use of this method, the calculations remain quite complex, so the automation is desirable.

In this paper we use the above mentioned program redesigned by V.S. to produce all four-loop vacuum supergraphs in ${\cal N}=1$ SQED with $N_f$ flavors regularized by higher derivatives. Using this result the four-loop beta-function is constructed in the form of an integral of double total derivatives. This integral is taken with the help of the program. The result is compared with three-loop anomalous dimension that was found in \cite{Shirokov:2022jyd}. As we will see below, the comparison demonstrates that NSVZ relation is really valid in four loops in the theory under consideration.

\section{${\cal N}=1$ SQED with $N_f$ flavors regularized by higher derivatives}

We consider massless ${\cal N}=1$ SQED with $N_f$ flavors
\begin{equation}
S = \frac{1}{4e_0^2} \operatorname{Re} \int d^4x\, d^2\theta\, W^a W_a + \frac{1}{4} \int d^4x\, d^4\theta\, \sum\limits_{\alpha=1}^{N_f} \Big(\phi_\alpha^* e^{2V} \phi_\alpha + \widetilde\phi_\alpha^* e^{-2V} \widetilde\phi_\alpha\Big),
\end{equation}

\noindent
which contains $N_f$ pairs of matter superfields $\phi_\alpha$ and $\widetilde\phi_\alpha$ with opposite charges with respect to the gauge group $U(1)$ and the gauge superfield $V$.

We use the higher derivatives regularization method \cite{Slavnov:1971aw,Slavnov:1972sq}, so we need to insert a special function of derivatives into the action,

\begin{equation}
S_{\operatorname{\scriptsize reg}} = \frac{1}{4e_0^2} \operatorname{Re} \int d^4x\, d^2\theta\, W^a R(\partial^2/\Lambda^2) W_a + \frac{1}{4} \int d^4x\, d^4\theta\, \sum\limits_{\alpha=1}^{N_f} \Big(\phi_\alpha^* e^{2V} \phi_\alpha + \widetilde\phi_\alpha^* e^{-2V} \widetilde\phi_\alpha\Big) ,
\end{equation}

\noindent
where the regulator function $R(x)$ grows at infinity and is equal to 1 at $x=0$. As usual, we need to introduce the gauge fixing term

\begin{equation}
S_{\operatorname{\scriptsize gf}} = -\frac{1}{32\xi_0 e_0^2} \int d^4x\, d^4\theta\,D^2 V K(\partial^2/\Lambda^2) \bar D^2 V,
\end{equation}

\noindent
where $\xi_0$ is the bare gauge parameter. Here we also introduce the regulator function $K(x)$ that has the same properties as $R(x)$.

However, the higher derivatives regularization cannot eliminate one-loop divergences. To do this, we have to insert into the generating functional the Pauli--Villars determinant \cite{Slavnov:1977zf}

\begin{equation}
\operatorname{Det}(PV,M)^{-1} = \int D\Phi\, D\widetilde\Phi\, \exp(iS_\Phi),
\end{equation}

\noindent
where the Pauli--Villars action is

\begin{equation}
S_\Phi = \frac{1}{4}\int d^4x\,d^4\theta\, \Big(\Phi^* e^{2V} \Phi + \widetilde\Phi^* e^{-2V} \widetilde\Phi\Big) + \Big(\frac{M}{2}\int d^4x\, d^2\theta\, \widetilde\Phi\, \Phi  +\operatorname{c.c.}\Big).
\end{equation}

\noindent
The ratio of the Pauli--Villars mass $M$ to the regularization parameter $\Lambda$ should not depend on the coupling constant,

\begin{equation}
a \equiv \frac{M}{\Lambda} \ne a(e_0).
\end{equation}

So, the generating functional takes the form

\begin{equation}
Z[\operatorname{sources}]= \int DV\,\Big(\prod\limits_{\alpha=1}^{N_f} D\phi_\alpha D\widetilde\phi_\alpha\Big)\, \operatorname{Det}(PV, M)^{N_f} \exp\Big(iS_{\operatorname{\scriptsize reg}} + iS_{\operatorname{\scriptsize gf}} + iS_{\operatorname{\scriptsize sources}}\Big)
\end{equation}

The considered theory is renormalizable, so that all that divergences can be absorbed into the renormalization of the coupling constant and superfields. Divergences are usually encoded in renormalization group functions. One must distinguish renormalization group functions defined in terms of the bare coupling constant \cite{Kataev:2013eta},

\begin{equation}\label{RGFs_Bare}
\beta(\alpha_0) = \frac{d\alpha_0}{d\ln\Lambda}\bigg|_{\alpha=\operatorname{\scriptsize const}}; \qquad \gamma(\alpha_0) = - \frac{d\ln Z}{d\ln\Lambda}\bigg|_{\alpha=\operatorname{\scriptsize const}},
\end{equation}

\noindent
from the ones defined in terms of the renormalized coupling constant,

\begin{equation}\label{RGFs_Renormalized}
\widetilde\beta(\alpha) = \frac{d\alpha}{d\ln\mu}\bigg|_{\alpha_0=\operatorname{\scriptsize const}}; \qquad \widetilde\gamma(\alpha) = \frac{d\ln Z}{d\ln\mu}\bigg|_{\alpha_0=\operatorname{\scriptsize const}},
\end{equation}

\noindent
where $\mu$ is a renormalization point and $Z$ is the same for all $\phi_\alpha$ and $\widetilde\phi_\alpha$: $\phi_{\alpha} = \sqrt{Z} \phi_{\alpha,R}$, $\widetilde\phi_{\alpha} = \sqrt{Z}\widetilde\phi_{\alpha,R}$.

If we use the higher derivative regularization, then renormalization group functions (\ref{RGFs_Bare}) satisfy \cite{Stepanyantz:2011jy,Stepanyantz:2014ima} the NSVZ equation \cite{Vainshtein:1986ja,Shifman:1985fi}

\begin{equation}\label{NSVZ}
\frac{\beta(\alpha_0)}{\alpha_0^2} = \frac{N_f}{\pi}\Big(1-\gamma(\alpha_0)\Big)
\end{equation}

\noindent
in all loops for an arbitrary renormalization prescription.

Renormalization group functions defined by Eq. (\ref{RGFs_Renormalized}) satisfy a similar equation only for specific renormalization prescriptions called ``the NSVZ schemes''. For example, some such schemes are given by HD+MSL renormalization prescription \cite{Kataev:2013eta, Shakhmanov:2017wji,Stepanyantz:2017sqg}.

\section{Method of calculating the $\beta$-function based on vacuum supergraph calculation}
\hspace*{\parindent}\label{VSC}
According to \cite{Stepanyantz:2019ihw}, there is a simple method for calculating the $\beta$-function in ${\cal N}=1$ supersymmetric theories. The standard procedure is based on calculation of supergraphs with two external gauge lines. Usually, it is rather complicated. The new method is based on calculating supergraphs with no external lines and inserting into the resulting loop integrals some specially constructed differential operators. Then for each vacuum supergraph we will obtain a contribution to the $\beta$-function which is equal to the one produced by all superdiagrams obtained by inserting two gauge lines into the original supergraph in all possible ways. This method was previously used in several calculations \cite{Stepanyantz:2019ihw,Kuzmichev:2019ywn,Stepanyantz:2019lyo, Aleshin:2020gec, Aleshin:2022zln}. In \cite{Aleshin:2020gec} this method was applied to ${\cal N}=1$ SQED with $N_f$ flavours regularized by higher derivatives. In this case we the calculation should be done according to the following prescription:

1. First, we calculate a supergraph with an insertion of $\theta^4 v^2$, where $v$ is function of the space-time coordinates which slowly decreases at a very large scale $L\to \infty$ and $\theta^4\equiv \theta^a \theta_a \bar\theta^{\dot b}\bar\theta_{\dot b}$.

2. Then we insert the operator

\begin{equation}\label{operator}
\sum\limits_{i=1}^M \frac{\partial^2}{\partial Q_{\mu i}^2},
\end{equation}

\noindent
into the corresponding loop integral. Here the index $i$ numerates the matter loops, $Q_{\mu i}$ is the Euclidean momentum of the $i$-th matter loop, and $M$ denotes a total number of matter loops.

3. Finally, the contribution to the function $\beta(\alpha_0)/\alpha_0^2$ produced by considered supergraph is obtained by differentiating the result with respect to $\ln\Lambda$ and multiplying it to the factor $-2\pi/{\cal V}_4$, where
\begin{equation}
{\cal V}_4 \equiv \int d^4x\, v^2.
\end{equation}

Note that the result is automatically obtained in the form of an integral of double total derivatives, and only singular contributions give non-trivial results. This can be seen with the help of the identity \cite{Bender:1976pw}
\begin{equation}\label{dd}
\frac{\partial^2}{\partial Q_{\mu }^2} \frac{1}{Q^2}=-4\pi^2 \delta^{4}(Q).
\end{equation}
Using it we obtain that the integral of double total derivatives can be calculated according to the rule \cite{Stepanyantz:2019ihw}
\begin{equation}\label{idd}
I=\int \frac{d^4 Q}{2\pi^4} \frac{\partial^2}{\partial Q_{\mu }^2}\bigg( \frac{1}{Q^2} f(Q^2) \bigg)=\frac{1}{4\pi^2}f(0)
\end{equation}
if $f$ is a nonsingular function rapidly falling at infinity. This in particular implies that all terms in which double total derivatives act on nonsingular expressions vanish.
\section{The four-loop $\beta$-function: Example}
\begin{figure}[h]
\hspace{6.5cm}
\includegraphics[scale=0.4]{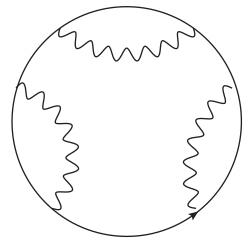}
\caption{Example superdiagram}
\label{ris:image}
\end{figure}

\hspace*{\parindent}
According to the previous section, in order to obtain a contribution to the four-loop $\beta$-function, we need to calculate a corresponding vacuum supergraph. Then we should insert the differential operator (\ref{operator}) into the resulting integrals. This can been done with the help of a special C++ program \cite{Shirokov:2022qdw} written previously by I.S. and adopted to calculation of the vacuum supergraphs by V.S. Let us illustrate the calculation process using the example of the superdiagram shown in Fig.~\ref{ris:image} in which we insert the expression $\theta^4 v^2$

For simplicity, we will do this in the Feynman gauge $\xi=1$. The program output for this specific superdiagram takes form:
\begin{eqnarray}\label{4Loop_Beta1}
&&\hspace*{-5mm}  -\frac{16 {\cal V}_4 N_f}{3} \int \frac{d^4 Q}{(2\pi)^4}
\frac{d^4K}{(2\pi)^4} \frac{d^4L}{(2\pi)^4} \frac{d^4U}{(2\pi)^4}\frac{e_0^6}{K^2 R_K L^2 R_L Q^2 R_Q} \bigg(\frac{1}{ U^2(U+L)^2(U+Q)^2(U+K)^2}
\nonumber\\
&&\hspace*{-5mm} -\frac{1}{ (U^2+M^2) ((U+K)^2+M^2)((U+L)^2+M^2)((U+Q)^2+M^2)}+\frac{5 M^2}{ (U+K)^2+M^2}
\nonumber\\
&&\hspace*{-5mm}\times \frac{1}{ (U^2+M^2)^2 ((U+L)^2+M^2)((U+Q)^2+M^2)}-\frac{4 M^4}{  ((U+K)^2+M^2)((U+L)^2+M^2)}\nonumber\\
&&\hspace*{-5mm}\times \frac{1}{ (U^2+M^2)^3 ((U+Q)^2+M^2) }\bigg)
\vphantom{\frac{1}{2}} ,
\end{eqnarray}
where $R_K \equiv R(K^2/\Lambda^2)$. After completing steps 2 and 3 from the algorithm described in the previous section, we obtain the contribution to the $\beta$-function
\begin{eqnarray}\label{4Loop_Beta2}
&&\hspace*{-5mm} \frac{\Delta\beta(\alpha_0)}{\alpha_0^2} = \frac{32 \pi N_f}{3} \frac{d}{d\ln\Lambda}\int \frac{d^4 Q}{(2\pi)^4}
\frac{d^4K}{(2\pi)^4} \frac{d^4L}{(2\pi)^4} \frac{d^4U}{(2\pi)^4}\frac{e_0^6}{K^2 R_K L^2 R_L Q^2 R_Q} \frac{\partial^2}{\partial U^\mu \partial U_\mu} \bigg(\frac{1}{ (U+K)^2}
\nonumber\\
&&\hspace*{-5mm} \times \frac{1}{U^2(U+L)^2(U+Q)^2} -\frac{1}{ (U^2+M^2) ((U+K)^2+M^2)((U+L)^2+M^2)((U+Q)^2+M^2)}
\nonumber\\
&&\hspace*{-5mm}+\frac{5 M^2}{ (U^2+M^2)^2 ((U+K)^2+M^2)((U+L)^2+M^2)((U+Q)^2+M^2)}-\frac{4 M^4}{  (U+K)^2+M^2}\nonumber\\
&&\hspace*{-5mm}\times \frac{1}{ (U^2+M^2)^3 ((U+L)^2+M^2)((U+Q)^2+M^2) }\bigg)
\vphantom{\frac{1}{2}} .
\end{eqnarray}

\begin{figure}[h]
\hspace{2.5cm}
\includegraphics[scale=0.4]{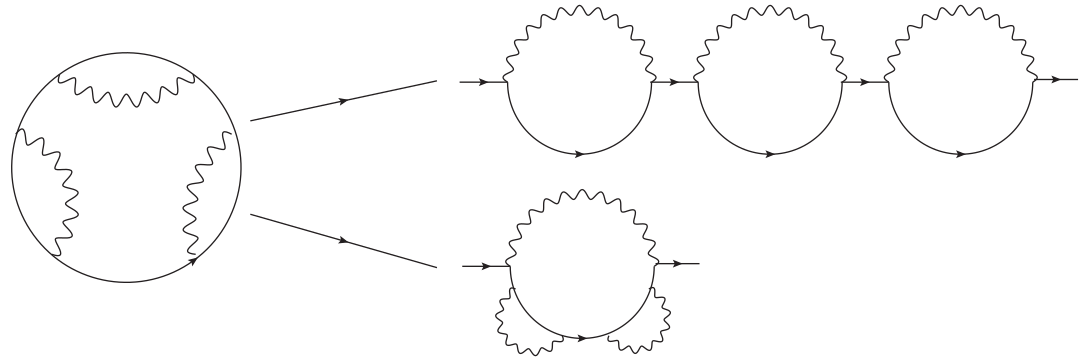}
\caption{Cutting of matter lines}
\label{ris:image1}
\end{figure}

Next, it is necessary to take integrals of double total derivatives using the equations analogous to (\ref{idd}). For some contributions this calculation is not trivial. That is why, a simple utility was created to take these integrals automatically. The result obtained with the help of it is written as
\begin{eqnarray}\label{Beta1}
 \frac{\Delta\beta(\alpha_0)}{\alpha_0^2} = \frac{8N_f}{\pi}\frac{d}{d\ln\Lambda} \int \frac{d^4K}{(2\pi)^4} \frac{d^4L}{(2\pi)^4} \frac{d^4Q}{(2\pi)^4} \frac{ e_0^6}{ K^4 R_K  L^2 R_L Q^2 R_Q}\bigg( \frac{1}{3 L^2 Q^2} + \frac{1}{  (K+L)^2 (K+Q)^2} \bigg).\nonumber \\
\vphantom{\frac{1}{2}}
\end{eqnarray}
According to Ref. \cite{Smilga:2004zr}, calculating the integral over $d^4U$ we should obtain certain contributions to the function $\ln G$, where $G$ is defined by following formula
\begin{eqnarray}\label{Two-Point_Functions}
&& \Gamma^{(2)} - S_{\mbox{\scriptsize gf}} = - \frac{1}{16\pi} \int \frac{d^4p}{(2\pi)^4}\, d^4\theta\, V(-p,\theta) \partial^2 \Pi_{1/2} V(p,\theta)\, d^{-1}(\alpha_0, \Lambda/p)\qquad\nonumber\\
&& + \frac{1}{4}\sum\limits_{\alpha=1}^{N_f} \int \frac{d^4q}{(2\pi)^4}\, d^4\theta\, \Big( \phi_\alpha^{*}(-q,\theta)\, \phi_\alpha(q,\theta) + \widetilde\phi_\alpha^{*}(-q,\theta)\, \widetilde\phi_\alpha(q,\theta)\Big)\, G(\alpha_0,\Lambda/q),
\end{eqnarray}
This can be graphically interpreted as cutting a certain matter line. In our case this process is illustrated by Fig.~\ref{ris:image1}
Here we have two contributions: the lower supergraph is 1PI and the upper one is not. The upper one corresponds to the third term in the series expansion of $\ln G = \Delta G - (\Delta G)^2/2 + (\Delta G)^3/3\ldots$. The expression for the one loop part of the upper graph can be found in \cite{Aleshin:2020gec} and in Feynman gauge has following form:
\begin{equation}\label{d1}
\Delta_{\mbox{\scriptsize 1}} G\Big|_{q=0} = - 2 e_0^2 \int \frac{d^4K}{(2\pi)^4} \frac{1}{K^4 R_K}.
\end{equation}
Contribution of the lower graph in Fig.~\ref{ris:image1} was calculated among others in \cite{Shirokov:2022jyd}. It is given by the expression:
\begin{equation}\label{d2}
\Delta_{\mbox{\scriptsize 2}} G\Big|_{q=0} = - 8 e_0^6 \int \frac{d^4K}{(2\pi)^4}\frac{d^4L}{(2\pi)^4} \frac{d^4Q}{(2\pi)^4} \frac{1}{ K^4 R_K  L^2 R_L Q^2 R_Q (K+L)^2 (K+Q)^2}.
\end{equation}
Comparing (\ref{Beta1}), (\ref{d1}) and (\ref{d2}) we see that they satisfy the equation
\begin{equation}\label{ns}
\frac{\Delta\beta(\alpha_0)}{\alpha_0^2}=- \frac{N_f}{\pi} \frac{d}{d\ln\Lambda} \Big(\Delta_{\mbox{\scriptsize 2}} G + \frac{1}{3} \big(\Delta_{\mbox{\scriptsize 1}} G\big)^3\Big)\Big|_{q=0}
\end{equation}
which exactly agrees with NSVZ relation
\begin{equation}\label{NSVZ2}
\frac{\beta(\alpha_0)}{\alpha_0^2} = \frac{N_f}{\pi}\Big(1-\frac{d\ln G}{d\ln\Lambda}\bigg|_{q=0}\Big)= \frac{N_f}{\pi}\Big(1-\gamma(\alpha_0)\Big).
\end{equation}

\section{The four-loop $\beta$-function: Full result and and the NSVZ relation}
\hspace*{\parindent}
To obtain full result we need to calculate all corresponding vacuum supergraphs. The initial program output and list of superdiagrams can be found in appendix \ref{Appendix_res}.

Next, it is necessary to take integrals of double total derivatives using the equations analogous to (\ref{idd}). Using utility mentioned in previous section we obtain result in the following form
\begin{eqnarray}\label{4-loop beta 3}
&&\hspace*{-5mm} \frac{\beta(\alpha_0)}{\alpha_0^2} =\frac{N_f}{\pi} +\frac{2N_f}{\pi} \frac{d}{d\ln\Lambda} \int \frac{d^4K}{(2\pi)^4} \frac{e_0^2}{K^4 R_K}
-\frac{2N_f}{\pi} \frac{d}{d\ln\Lambda} \int \frac{d^4K}{(2\pi)^4} \frac{d^4L}{(2\pi)^4} \frac{e_0^4}{R_K R_L} \bigg(\frac{2}{K^2 L^4 (K+L)^2}\nonumber\\
&&\hspace*{-5mm} - \frac{1}{K^4 L^4} \bigg)
-\frac{4(N_f)^2}{\pi} \frac{d}{d\ln\Lambda} \int \frac{d^4K}{(2\pi)^4} \frac{d^4L}{(2\pi)^4} \frac{e_0^4}{R_K^2 K^4} \bigg(\frac{1}{L^2 (L+K)^2} - \frac{1}{(L^2+M^2)((L+K)^2+M^2)}\bigg)\nonumber\\
&&\hspace*{-5mm} - \frac{8N_f}{\pi}\frac{d}{d\ln\Lambda} \int \frac{d^4K}{(2\pi)^4} \frac{d^4L}{(2\pi)^4} \frac{d^4Q}{(2\pi)^4} \frac{ e_0^6}{R_K R_L R_Q}\bigg[- \frac{1}{3K^4 L^4 Q^4} + \frac{1}{K^4 L^2 Q^4 (Q+L)^2} + \frac{1}{K^2  (K+L)^2}\nonumber\\
&&\hspace*{-5mm} \times \frac{1}{L^4(Q+K+L)^2}\bigg(\frac{1}{Q^2} - \frac{2}{(Q+L)^2}\bigg) \bigg]
-\frac{16(N_f)^2}{\pi} \frac{d}{d\ln\Lambda} \int \frac{d^4K}{(2\pi)^4} \frac{d^4L}{(2\pi)^4} \frac{d^4Q}{(2\pi)^4} \frac{e_0^6\, K_\mu L^\mu}{R_K^2 R_L  (K+L)^2}\nonumber\\
&&\hspace*{-5mm} \times \frac{1}{K^4L^4} \bigg(\frac{1}{Q^2 (Q+K)^2} - \frac{1}{(Q^2+M^2)((Q+K)^2+M^2)}\bigg)
-\frac{8(N_f)^2}{\pi} \frac{d}{d\ln\Lambda} \int \frac{d^4K}{(2\pi)^4} \frac{d^4L}{(2\pi)^4} \frac{d^4Q}{(2\pi)^4} \nonumber\\
&&\hspace*{-5mm} \times \frac{e_0^6}{R_K^2 R_L K^4 L^2}\bigg(\frac{2(Q+K+L)^2-K^2-L^2}{Q^2(Q+K)^2(Q+L)^2(Q+K+L)^2}
- \frac{2(Q+K+L)^2 - K^2 - L^2}{((Q+K)^2+M^2)((Q+L)^2+M^2)}\nonumber\\
&&\hspace*{-5mm} \times \frac{1}{(Q^2+M^2)((Q+K+L)^2+M^2)} + \frac{4M^2}{(Q^2+M^2)^2 ((Q+K)^2+M^2) ((Q+L)^2+M^2)}
\bigg)
 \nonumber\\
&&\hspace*{-5mm}+\frac{8(N_f)^3}{\pi} \frac{d}{d\ln\Lambda} \int \frac{d^4K}{(2\pi)^4} \frac{d^4L}{(2\pi)^4} \frac{d^4Q}{(2\pi)^4} \frac{ e_0^6}{R_K^3 K^4}\bigg(\frac{1}{Q^2 (Q+K)^2} - \frac{1}{(Q^2+M^2)((Q+K)^2 +M^2)}\bigg) \nonumber\\
&&\hspace*{-5mm} \times \bigg(\frac{1}{L^2 (L+K)^2}- \frac{1}{(L^2+M^2)((L+K)^2 +M^2)}\bigg) + O(e_0^8).
\end{eqnarray}
All integrals that remain in this expression were taken in \cite{Shirokov:2022jyd} with the help of the Chebyshev polynomials method \cite{Rosner:1967zz}. So, using the corresponding results we can present the four-loop $\beta$-function defined in terms of the bare coupling constant in the form
\begin{eqnarray}
&& \beta(\alpha_0) = \frac{\alpha_0^2 N_f}{\pi} + \frac{\alpha_0^3 N_f}{\pi^2} - \frac{\alpha_0^4 N_f}{2\pi^3} - \frac{\alpha_0^4 (N_f)^2}{\pi^3}\Big(\ln a + 1 + \frac{A_1}{2}\Big) + \frac{\alpha_0^5 N_f}{2\pi^4}
+ \frac{\alpha_0^5 (N_f)^2}{\pi^4} \qquad\nonumber\\
&& \times \Big(\ln a + \frac{3}{4} + C\Big) + \frac{\alpha_0^5(N_f)^3}{\pi^4} \Big((\ln a + 1)^2 -\frac{A_2}{4} + D_1 \ln a + D_2\Big) + O(\alpha_0^6),\qquad
\end{eqnarray}
where
\begin{eqnarray}\label{Const}
&& \hspace*{-7mm} A_1 \equiv \int\limits_0^\infty dx\, \ln x\, \frac{d}{dx}\Big(\frac{1}{R(x)}\Big);\qquad\qquad\qquad\qquad\qquad
A_2 \equiv \int\limits_0^\infty dx\, \ln^2 x\, \frac{d}{dx}\Big(\frac{1}{R(x)}\Big);\nonumber\\
&& \hspace*{-7mm} C \equiv \int\limits_0^1 dx \int\limits_0^\infty dy\, x\, \ln y\, \frac{d}{dy}\Big(\frac{1}{R(y) R(x^2 y)}\Big);\qquad\qquad\ D_1 \equiv \int\limits_0^\infty dx\, \ln x\, \frac{d}{dx}\Big(\frac{1}{R^2(x)}\Big);\nonumber\\
&& \hspace*{-7mm} D_2 \equiv \int\limits_0^\infty dx\, \ln x\, \frac{d}{dx}\left\{\frac{1}{R^2(x)}\bigg[-\frac{1}{2}\Big(1-R(x)\Big) \ln x + \sqrt{1+\frac{4a^2}{x}}\, \operatorname{arctanh}\sqrt{\frac{x}{x+4a^2}}\,\bigg]\right\}
\end{eqnarray}
are the constants which depend on a particular form of $R(x)$. Note that lower orders here were found in \cite{Aleshin:2020gec} and were independently checked in this research.

The three-loop anomalous dimension of the matter superfields was found in \cite{Shirokov:2022jyd}. It can be written as
\begin{eqnarray}\label{Gamma_3}
&& \gamma(\alpha_0) = - \frac{\alpha_0}{\pi} + \frac{\alpha_0^2}{2\pi^2} + \frac{\alpha_0^2 N_f}{\pi^2}\Big(\ln a + 1 + \frac{A_1}{2}\Big) - \frac{\alpha_0^3}{2\pi^3} - \frac{\alpha_0^3 N_f}{\pi^3}\Big(\ln a +\frac{3}{4} +C\Big)\nonumber\\
&& - \frac{\alpha_0^3(N_f)^2}{\pi^3} \Big((\ln a + 1)^2 -\frac{A_2}{4} + D_1 \ln a + D_2\Big) + O(\alpha_0^4).
\end{eqnarray}

Comparing the results for the four-loop $\beta$-function and for the three-loop anomalous dimension we see that they are really related to each other by the NSVZ relation (\ref{NSVZ}).

For completeness, let us also present the expressions for renormalization group functions defined in terms of renormalized coupling constant.

\begin{eqnarray}\label{Beta4}
&&\hspace*{-5mm} \frac{\widetilde\beta(\alpha)}{\alpha^2} = -\frac{d}{d\ln\mu}\Big(\frac{1}{\alpha}\Big)\bigg|_{\alpha_0 = \mbox{\scriptsize const}} = \frac{N_f}{\pi} + \frac{\alpha N_f}{\pi^2} - \frac{\alpha^2 N_f}{2\pi^3} - \frac{\alpha^2 (N_f)^2}{\pi^3}\Big(\ln a + 1+\frac{A_1}{2} + b_{2,0} - b_{1,0}\Big)
\nonumber\\
&&\hspace*{-5mm} + \frac{\alpha^3 N_f}{2\pi^4} + \frac{\alpha^3 (N_f)^2}{\pi^4} \Big(\ln a + \frac{3}{4} + C + b_{3,0} - b_{1,0}\Big) + \frac{\alpha^3 (N_f)^3}{\pi^4} \Big\{ \Big(\ln a +1 - b_{1,0}\Big)^2 - b_{1,0} A_1
+ b_{3,1} \nonumber\\
&&\hspace*{-5mm}  -\frac{A_2}{4} + D_1 \ln a + D_2   \Big\} + O(\alpha^4).\vphantom{\frac{1}{2}}
\end{eqnarray}
\begin{eqnarray}\label{Gamma3}
&&\hspace*{-5mm} \widetilde\gamma(\alpha)= - \frac{\alpha}{\pi} + \frac{\alpha^2}{2\pi^2} + \frac{\alpha^2 N_f}{\pi^2}\Big(\ln a + 1 + \frac{A_1}{2} + g_{1,0} - b_{1,0}\Big) - \frac{\alpha^3}{2\pi^3}
+ \frac{\alpha^3 N_f}{\pi^3} \Big(- \ln a - \frac{3}{4} - C   \nonumber\\
&&\hspace*{-5mm} - b_{2,0} + b_{1,0} - g_{2,0} + g_{1,0}\Big) + \frac{\alpha^3 (N_f)^2}{\pi^3}\Big\{-\Big(\ln a + 1 - b_{1,0}\Big)^2 + \frac{A_2}{4} - D_1 \ln a - D_2 + b_{1,0} A_1\nonumber\\
&&\hspace*{-5mm} - g_{2,1}\Big\} + O(\alpha^4). \vphantom{\frac{N_f^2}{\pi^2}}
\end{eqnarray}
where the constants $b_{1,0}, b_{2,0}, b_{3,0}, b_{3,1}, g_{1,0}, g_{2,0}$ fix a renormalization scheme in the lowest orders. A detailed study of different renormalization schemes in this case can be found in \cite{Shirokov:2022jyd}.

\section{Conclusion}
\hspace*{\parindent}

In this paper we have directly calculated the four-loop $\beta$-function for ${\cal N}=1$ SQED with $N_f$ flavors regularized by higher derivatives in the general $\xi$-gauge. This was done using a special method based on calculating vacuum supergraphs and the computer program originally written by I.S. \cite{Shirokov:2022qdw}. This program was modified by V.S. in order to make it suitable for dealing with vacuum supergraphs. The result was obtained in the form of integrals of total double derivatives. Using a special utility these integrals were taken. The result exactly coincided with the expression obtained in \cite{Shirokov:2022jyd} with the help of the NSVZ relation. This in particular implies that the four-loop $\beta$-function defined in terms of bare couplings is really related to the three-loop anomalous dimension (directly calculated in \cite{Shirokov:2022jyd}). Therefore, we have explicitly demonstrated that NSVZ relation for renormalization group functions defined in terms of bare couplings is really valid in the four-loop approximation for this theory, thus confirming the correctness of the method and software used for calculating the $\beta$-function and the general results about the validity of the NSVZ equation in theories regularized by higher (covariant) derivatives.

\section*{Acknowledgments}
\hspace*{\parindent}

The authors are very grateful to K.V.Stepanyantz to his valuable discussions and useful comments on the manuscript. The authors also thank A.L. Kataev for his important remarks. All the graphs were drawn using JaxoDraw software \cite{Binosi:2003yf}.

The work was supported by Foundation for Advancement of Theoretical Physics and Mathematics ``BASIS'', grant  No. 23-1-4-28-1 (I.S.).

\appendix

\section{Contributions to 4-loop $\beta$-function}
\begin{figure}[h!]
\hspace{-0.1cm}
\includegraphics[scale=0.18]{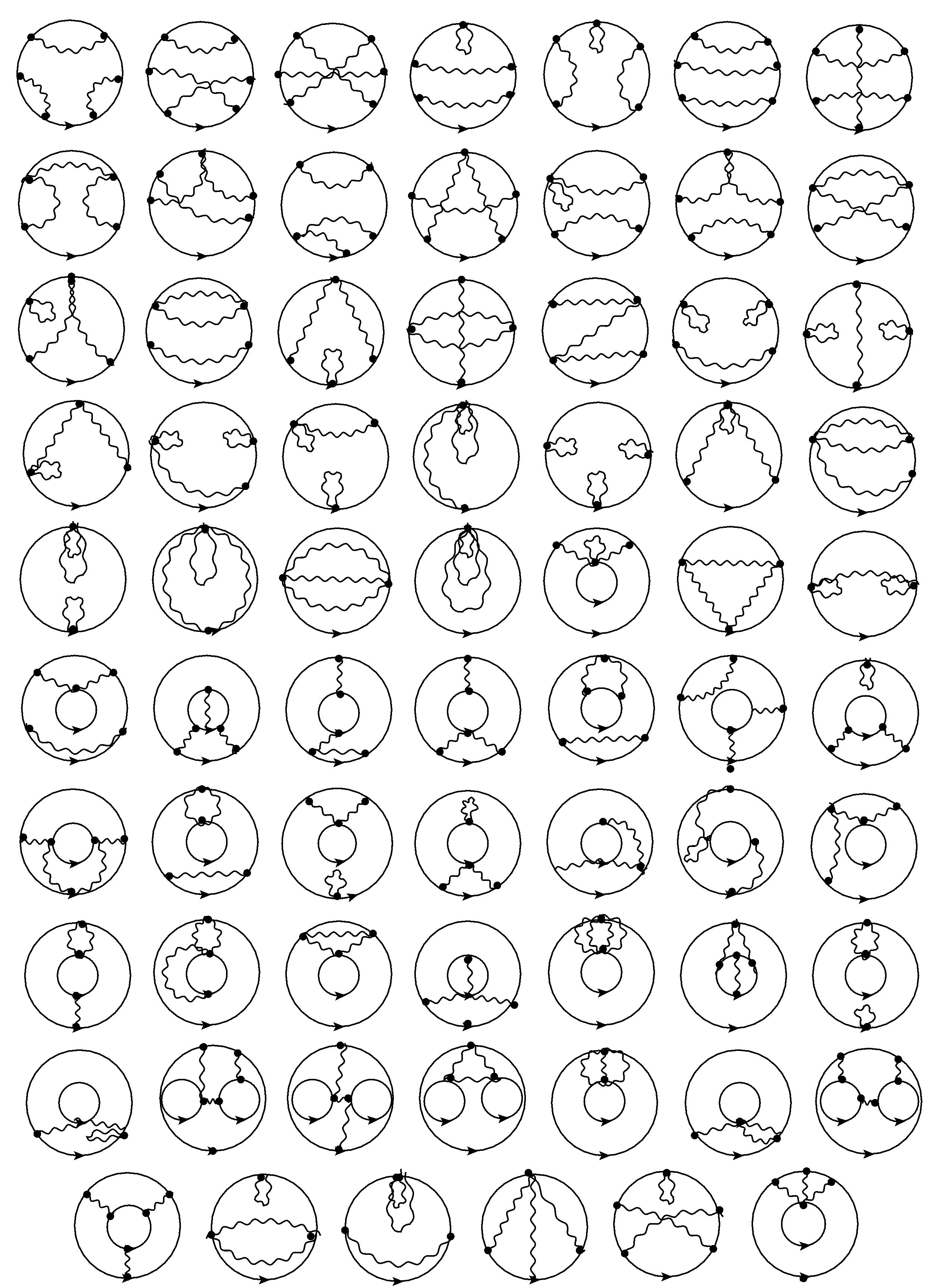}
\caption{Vacuum supergraphs contributing to 4-loop $\beta$-function}
\label{ris:image3}
\end{figure}

\hspace*{\parindent}\label{Appendix_res}
Here we present initial program output for all the graphs from Fig.~\ref{ris:image3}.

It is convenient to split the result into several parts that contain different powers of $N_f$. (Note that the degree of $N_f$ is equal to the number of matter loops and to the number of terms in the operator (\ref{operator}).) The result obtained after making all steps described in the previous section can be written in the form
\begin{eqnarray}\label{4Loop_Beta}
&&\hspace*{-5mm} \frac{\beta(\alpha_0)}{\alpha_0^2} = 2\pi N_f \frac{d}{d\ln\Lambda}
\int \frac{d^4Q}{(2\pi)^4} \frac{\partial^2}{\partial Q^\mu \partial Q_\mu}
\frac{\ln(Q^2+M^2)}{Q^2} + 4\pi N_f \frac{d}{d\ln\Lambda} \int \frac{d^4Q}{(2\pi)^4} \frac{d^4K}{(2\pi)^4} \frac{e_0^2}{K^2 R_K}
\nonumber\\
&&\hspace*{-5mm} \times  \frac{\partial^2}{\partial Q^\mu \partial Q_\mu} \bigg(\frac{1}{Q^2 (Q+K)^2} -\frac{1}{(Q^2+M^2)( (Q+K)^2+M^2)}\bigg)+ 8 \pi (N_f)^2 \frac{d}{d\ln\Lambda} \int \frac{d^4Q}{(2\pi)^4} \frac{d^4K}{(2\pi)^4} \nonumber\\
&&\hspace*{-5mm}  \times \frac{d^4L}{(2\pi)^4} \frac{e_0^4}{K^2 R_K^2}\frac{\partial^2}{\partial Q^\mu \partial Q_\mu} \bigg(\frac{1}{Q^2 (Q+K)^2} -\frac{1}{(Q^2+M^2)( (Q+K)^2+M^2)}\bigg) \bigg(\frac{1}{
((L+K)^2+M^2)}\nonumber\\
&&\hspace*{-5mm} \times  \frac{1}{(L^2+M^2)} -\frac{1}{L^2 (L+K)^2}  \bigg)\bigg] + 8\pi N_f \frac{d}{d\ln\Lambda} \int \frac{d^4Q}{(2\pi)^4}\frac{d^4K}{(2\pi)^4} \frac{d^4L}{(2\pi)^4} \frac{e_0^4}{K^2 R_K L^2 R_L} \frac{\partial^2}{\partial Q^\mu \partial Q_\mu}\bigg( \frac{1}{Q^2}
\nonumber\\
&&\hspace*{-5mm} \times  \frac{1}{(Q+K)^2(Q+L)^2}- \frac{K^2}{Q^2 (Q+K)^2 (Q+L)^2(Q+K+L)^2} +\frac{K^2 + M^2}{(Q^2+M^2)((Q+L)^2+M^2)} \nonumber\\
&&\hspace*{-5mm} \times \frac{1}{((Q+K)^2+M^2)((Q+K+L)^2+M^2)} - \frac{1}{(Q^2+M^2) ((Q+K)^2+M^2)((Q+L)^2+M^2)} \nonumber\\
&&\hspace*{-5mm} +\frac{2M^2}{(Q^2+M^2)^2((Q+K)^2+M^2) ((Q+L)^2+M^2)} \bigg) + 8\pi N_f \frac{d}{d\ln\Lambda}\int \frac{d^4 Q}{(2\pi)^4}
\frac{d^4K}{(2\pi)^4} \frac{d^4L}{(2\pi)^4} \frac{d^4U}{(2\pi)^4}
\nonumber\\
&&\hspace*{-5mm} \times \frac{e_0^6}{K^2 R_K L^2 R_L Q^2 R_Q} \frac{\partial^2}{\partial U^\mu \partial U_\mu} \bigg( f(K,Q,L,U,M)-\frac{3}{ U^2 (U+K)^2(U+K+L)^2(U+Q)^2}-\frac{4}{ U^2 }
\nonumber\\
&&\hspace*{-5mm} \times \frac{1}{ (U+L)^2(U+K+L)^2(U+Q+K)^2 }+\frac{2L^2 Q^2}{ U^2(U+L)^2 (U+K)^2(U+K+L)^2(U+Q+L)^2 }
\nonumber\\
&&\hspace*{-5mm} \times \frac{1}{ (U+Q)^2 }-\frac{1}{U^2(U+L)^2 (U+Q)^2 (U+K+L)^2}-\frac{4L^2}{(U+Q)^2(U+L)^2(U+K)^2 (U+K+L)^2}
\nonumber\\
&&\hspace*{-5mm} \times \frac{1}{ U^2 }+\frac{(K+Q)^2 \times[3(U+L+K)^2-(U+K)^2-(U+L)^2+U^2+L^2]}{U^2(U+L)^2(U+K)^2 (U+K+Q+L)^2 (U+K+L)^2 (U+Q+K)^2}+\frac{2K^4}{U^2(U+L)^2}
\nonumber\\
&&\hspace*{-5mm} \times\frac{1}{(U+K)^2 (U+Q)^2 (U+K+L)^2 (U+Q+K)^2}  +\frac{2(Q+K+L)^4}{3U^2(U+L)^2(U+K)^2 (U+K+Q+L)^2}
\nonumber\\
&&\hspace*{-5mm} \times\frac{1}{(U+Q+L)^2 (U+Q+K)^2}+\frac{16[(U+K)^2+(U+L)^2+(U+Q)^2-U^2-L^2-K^2-Q^2]}{3U^2(U+L)^2(U+K)^2 (U+Q+L)^2  (U+Q+K)^2}
\nonumber\\
&&\hspace*{-5mm} +\frac{4}{3 U^2 (U+K)^2(U+L)^2(U+Q)^2}\bigg)+16\pi (N_f)^2  \frac{d}{d\ln\Lambda}\int \frac{d^4 Q}{(2\pi)^4}
\frac{d^4K}{(2\pi)^4} \frac{d^4L}{(2\pi)^4} \frac{d^4U}{(2\pi)^4}\frac{e_0^6}{ R_K^2 K^2 R_L L^2}
\nonumber\\
&&\hspace*{-5mm} \times \bigg(\frac{\partial^2}{\partial Q^\mu \partial Q_\mu}+\frac{\partial^2}{\partial U^\mu \partial U_\mu}\bigg)\bigg( \frac{K^2+L^2-2(Q+K+L)^2}{U^2(U+K)^2Q^2(Q+K)^2(Q+L)^2(Q+K+L)^2}-\frac{1}{(U+K)^2+M^2}
\nonumber\\
&&\hspace*{-5mm} \times \frac{K^2+L^2-2(Q+K+L)^2}{(U^2+M^2)(Q+K)^2 Q^2(Q+L)^2(Q+K+L)^2) }-\frac{K^2+L^2-2(Q+K+L)^2}{U^2(U+K)^2(Q^2+M^2)((Q+K)^2+M^2)}
\nonumber\\
&&\hspace*{-5mm} \times \frac{1}{((Q+L)^2+M^2)((Q+K+L)^2+M^2)}-\frac{4M^2}{U^2(Q^2+M^2)^2((Q+K)^2+M^2)((Q+L)^2+M^2)}
\nonumber\\
&&\hspace*{-5mm}\times \frac{1}{(U+K)^2}-\frac{K^2+L^2-2(Q+K+L)^2}{(U^2+M^2)(Q^2+M^2)((Q+K)^2+M^2)((Q+L)^2+M^2)((Q+K+L)^2+M^2)}
\nonumber\\
&&\hspace*{-5mm}\times \frac{1}{(U+K)^2+M^2}-\frac{4M^2}{((U+K)^2+M^2)(Q^2+M^2)^2((Q+K)^2+M^2)((Q+L)^2+M^2)}
\nonumber\\
&&\hspace*{-5mm}\times \frac{1}{U^2+M^2}  \bigg)+16\pi (N_f)^3  \frac{d}{d\ln\Lambda}\int \frac{d^4 Q}{(2\pi)^4}
\frac{d^4K}{(2\pi)^4} \frac{d^4L}{(2\pi)^4} \frac{d^4U}{(2\pi)^4}\frac{e_0^6}{ R_K^3 K^2}\frac{\partial^2}{\partial U^\mu \partial U_\mu}\bigg(\frac{1}{U^2 (U+K)^2}
\nonumber\\
&&\hspace*{-5mm}  - \frac{1}{((U+K)^2+M^2)(U^2+M^2)
} \bigg) \bigg(  \frac{1}{L^2(L+K)^2} - \frac{1}{(L^2+M^2)
((L+K)^2+M^2)} \bigg)\bigg(\frac{1}{Q^2 (Q+K)^2}
\nonumber\\
&&\hspace*{-5mm} - \frac{1}{((Q+K)^2+M^2)(Q^2+M^2)
}  \bigg)  + O(e_0^8)
\vphantom{\frac{1}{2}} ,
\end{eqnarray}
where $R_K \equiv R(K^2/\Lambda^2)$. The function $ f(K,Q,L,U,M)$ is presented in the appendix \ref{Appendix_nsc}. It is composed from all terms that contain only propagators of Pauli-Villars fields. These propagators have no singularities and, therefore, do not contribute to the result. Note that in the expression (\ref{4Loop_Beta}) we present not only the answer obtained after calculation of the four-loop vacuum supergraphs but also the contributions corresponding to  the lower orders. They were found in \cite{Aleshin:2020gec}, but for testing the software we have also recalculated them independently. The result coincided with the one obtained in \cite{Aleshin:2020gec}. Note that all terms proportional to $(N_f)^2$ come from the supegraphs containing the two-loop polarization operator, while all other contributions vanish due to the opposite signs of the $U(1)$ charges of the superfields $\phi_\alpha$ and $\widetilde\phi_\alpha$. The result was obtained in general $\xi$ gauge, but the gauge dependence disappeared as it must be according to \cite{Batalin:2019wkb}.
\section{Non-singular contributions present in the term proportional to $N_{f}$}
\hspace*{\parindent}\label{Appendix_nsc}
Here we wite down the contribution of the Pauli-Villars fields present in the term proportional to $N_f$. In (\ref{4Loop_Beta}) they are denoted by function $ f(K,Q,L,U,M)$ that is given by the expression
\begin{eqnarray}\label{f1}
&&\hspace*{-5mm} f(K,Q,L,U,M)= \frac{4}{(U^2+M^2) ((U+L)^2+M^2)((U+K+L)^2+M^2)((U+Q+K)^2+M^2) }
\nonumber\\
&&\hspace*{-5mm}+  \frac{3}{ (U^2+M^2) ((U+K)^2+M^2)((U+K+L)^2+M^2)((U+Q)^2+M^2)}-\frac{2L^2 Q^2}{ (U+K+L)^2+M^2}
\nonumber\\
&&\hspace*{-5mm} \times \frac{1}{((U+L)^2+M^2) ((U+K)^2+M^2)(U^2+M^2)((U+Q+L)^2+M^2)  ((U+Q)^2+M^2) }
\nonumber\\
&&\hspace*{-5mm} + \frac{1}{(U^2+M^2)((U+L)^2+M^2) ((U+Q)^2+M^2) ((U+K+L)^2+M^2)} +\frac{4L^2}{ (U+K+L)^2+M^2 }
\nonumber\\
&&\hspace*{-5mm} \times \frac{1}{ (U^2+M^2)((U+L)^2+M^2)((U+Q)^2+M^2) ((U+K)^2+M^2) }-\frac{1}{(U+K+Q+L)^2+M^2}\nonumber\\
&&\hspace*{-5mm}\times \frac{(K+Q)^2 \times[3(U+L+K)^2-(U+K)^2-(U+L)^2+U^2+L^2]}{(U^2+M^2) ((U+L)^2+M^2)((U+K)^2+M^2) ((U+K+L)^2+M^2) ((U+Q+K)^2+M^2)}\nonumber\\
&&\hspace*{-5mm}-\frac{2K^4}{(U^2+M^2)((U+L)^2+M^2)+M^2)((U+K)^2+M^2) ((U+Q)^2+M^2) ((U+K+L)^2+M^2) }
\nonumber\\
&&\hspace*{-5mm} \times\frac{1}{ (U+Q+K)^2+M^2}   -\frac{2(Q+K+L)^4}{3((U+L)^2+M^2)((U+K)^2+M^2) ((U+K+Q+L)^2+M^2)}
\nonumber\\
&&\hspace*{-5mm}  \times\frac{1}{( U^2+M^2)((U+Q+L)^2+M^2) ((U+Q+K)^2+M^2)} -\frac{1}{(U^2+M^2) ((U+Q+L)^2+M^2)}
\nonumber\\
&&\hspace*{-5mm}  \times \frac{16[(U+K)^2+(U+L)^2+(U+Q)^2-U^2-L^2-K^2-Q^2]}{3((U+L)^2+M^2)((U+K)^2+M^2) ((U+Q+K)^2+M^2)} -\frac{4}{3 (U^2+M^2) ((U+K)^2+M^2) }
\nonumber\\
&&\hspace*{-5mm}  \times \frac{1}{((U+L)^2+M^2)((U+Q)^2+M^2)} -\frac{4 M^2 (K+L)^2}{3 (U^2+M^2) ((U+L)^2+M^2)((U + K+L+Q)^2+M^2) }
\nonumber\\
&&\hspace*{-5mm}
\times \frac{1}{((U+Q)^2+M^2((U+K+L)^2+M^2)((U+Q+K)^2+M^2)} -\frac{4 M^2[ 4K^2+7(U+Q+L)^2]}{3 ((U+Q+L)^2+M^2)}
\nonumber\\
&&\hspace*{-5mm}
\times \frac{1}{((U+L)^2+M^2)((U+Q)^2+M^2)((U+K+L)^2+M^2)((U+Q+K)^2+M^2)}
\nonumber\\
&&\hspace*{-5mm}
\times \frac{1}{((U+K)^2+M^2)} +\frac{2 M^2[ Q^2-2K^2+6(U+K)^2+2(U+K+Q)^2+U^2]}{ (U^2+M^2) ((U+L)^2+M^2)((U+Q)^2+M^2) ((U+Q+K)^2+M^2)}
\nonumber\\
&&\hspace*{-5mm}
\times \frac{1}{((U+K)^2+M^2)((U+K+L)^2+M^2)}-\frac{2 M^2[2(K+Q)^2+(K+Q+L)^2+(L+Q)^2]}{ (U^2+M^2) ((U+L)^2+M^2) ((U+Q+K)^2+M^2)}
\nonumber\\
&&\hspace*{-5mm}
\times \frac{1}{((U+K)^2+M^2)((U+K+L)^2+M^2)((U+Q+K+L)^2+M^2)}+\frac{2 M^2(U+Q)^2}{(U+Q+K)^2+M^2 }
\nonumber\\
&&\hspace*{-5mm}
\times \frac{1}{((U+K)^2+M^2) ((U+L)^2+M^2)((U+Q+L)^2+M^2)(U^2+M^2) ((U+Q+K+L)^2+M^2)}
\nonumber\\
&&\hspace*{-5mm}
-\frac{8 M^2(K+L)^2}{ (U^2+M^2)^2 ((U+L)^2+M^2) ((U+Q)^2+M^2)((U+K)^2+M^2)((U+K+L)^2+M^2)}
\nonumber\\
&&\hspace*{-5mm}
+\frac{32 M^2}{ 3(U^2+M^2)^2 ((U+L)^2+M^2) ((U+Q)^2+M^2)((U+K)^2+M^2)}+\frac{9 M^4}{(U+K)^2+M^2}
\nonumber\\
&&\hspace*{-5mm}
\times \frac{1}{ (U^2+M^2) ((U+L)^2+M^2) ((U+Q)^2+M^2)((U+K+Q)^2+M^2)((U+K+L)^2+M^2)}
\nonumber\\
&&\hspace*{-5mm}
-\frac{10 M^4}{ ((U+Q)^2+M^2)((U+K+Q)^2+M^2)((U+L+Q)^2+M^2)((U+K+L)^2+M^2)}
\nonumber\\
&&\hspace*{-5mm}
\times \frac{1}{ ((U+K)^2+M^2) ((U+L)^2+M^2)}-\frac{16 M^4}{ (U^2+M^2)^2((U+K)^2+M^2)((U+K+L)^2+M^2)}
\nonumber\\
&&\hspace*{-5mm}
\times \frac{1}{ ((U+L)^2+M^2) ((U+Q)^2+M^2)}-\frac{8 M^4}{ (U^2+M^2)^2((U+K)^2+M^2)^2((U+K+Q)^2+M^2)}
\nonumber\\
&&\hspace*{-5mm}
\times \frac{1}{ (U+L)^2+M^2}-\frac{16 M^4}{3 (U^2+M^2)^3((U+K)^2+M^2)^2((U+L)^2+M^2)((U+Q)^2+M^2)}
\vphantom{\frac{1}{2}} .
\end{eqnarray}
To verify this expression we note that if $M$ is set to zero, then we obtain the term proportional to $N_f$ from  (\ref{4Loop_Beta}) with the opposite sign.

\end{document}